\documentclass[
 reprint,
 amsmath,amssymb,
 aps,
 prl,
]{revtex4-1}

\usepackage{graphicx}
\usepackage{dcolumn}
\usepackage{bm}
\usepackage{color}

\begin{document}

\title{Hole-flux Composite Fermion Commensurability Oscillations}

\author{D.\ Kamburov}
\author{M.\ Shayegan}
\author{L.N.\ Pfeiffer}
\author{K.W.\ West}
\author{K.W.\ Baldwin}
\affiliation{
Department of Electrical Engineering, Princeton University, Princeton, New Jersey 08544, USA
}
\date{\today}

\begin{abstract}
We report the observation of commensurability oscillations of hole-flux composite fermions near filling factor $\nu=1/2$ in a high-mobility two-dimensional hole system confined to a GaAs quantum well, and subjected to a weak, strain-induced, unidirectional periodic potential modulation. The oscillations, which are consistent with ballistic transport of fully spin-polarized composite fermions in a weak periodic effective magnetic field, are surprisingly strong and exhibit up to third-order minima. We extract a ballistic mean-free-path of about 0.2 $\mu$m for the hole-flux composite fermions.

\end{abstract}


\maketitle

In the presence of a strong perpendicular magnetic field and at very low temperatures, two-dimensional (2D) interacting carriers minimize their energy by capturing an even number of flux quanta and creating new particles called composite fermions (CFs) \cite{Jain.Composite.Fermions,Jain.PRL.63.199.1989,Halperin.PRB.47.7312.1993}.
When a Landau level (LL) is exactly half-filled, e.g. at filling factor $\nu=1/2$, the flux attachment cancels the external magnetic field $B=B_{1/2}$ completely, leaving the CFs in zero effective field. At and near half fillings the CFs form a Fermi sea with its own Fermi contour, and behave analogously to their zero-field counterparts in an effective field given by $B^{*}=B-B_{1/2}$. The quasi-classical orbits of CFs are thus expected to exhibit geometric resonances near $\nu=1/2$ when their size equals the length scale of an external potential disturbance, similar to what is seen at low magnetic fields. Such resonances were indeed observed for CFs and provided a direct confirmation of the validity of the CF picture \cite{Willett.PRL.71.3846.1993,Kang.PRL.71.3850.1993,Goldman.PRL.72.13.1994,Smet.PRL.77.2272.1996}.

CF Fermi contours were probed in a number of later experiments in which a small periodic potential modulation was introduced in high-quality 2D electron systems \cite{Willett.PRL.78.4478.1997,Smet.PRB.56.3598.1997,Smet.PRL.80.20.1998,Smet.PRL.83.2620.1999}. In such experiments, one expects oscillations in the magnetoresistance, signalling the commensurability of the CF quasi-classical cyclotron orbit diameter with multiple integers of the period of the potential modulation. The observed magnetoresistance features around $\nu=1/2$, however, were not clear oscillations but rather two weak minima, one on each side of $\nu=1/2$, corresponding to the first instance when the CF orbits became commensurate with the period of the modulation \cite{Willett.PRL.78.4478.1997,Smet.PRB.56.3598.1997,Smet.PRL.80.20.1998,Smet.PRL.83.2620.1999,Zwerschke.PRL.83.2616.1999,Peeters.PRB.47.1466.1993,Gerhardts.PRB.53.11064.1996,Oppen.PRL.80.4494.1998,Mirlin.PRL.81.1070.1998}. More recent experiments have hinted at additional magnetoresistance minima near $\nu=3/2$ although their positions and amplitudes are not entirely consistent with simple commensurability oscillations (COs) of CFs \cite{Endo.PRB.63.113310.2001}.

\begin{figure}[h!]
\includegraphics[width=0.48\textwidth]{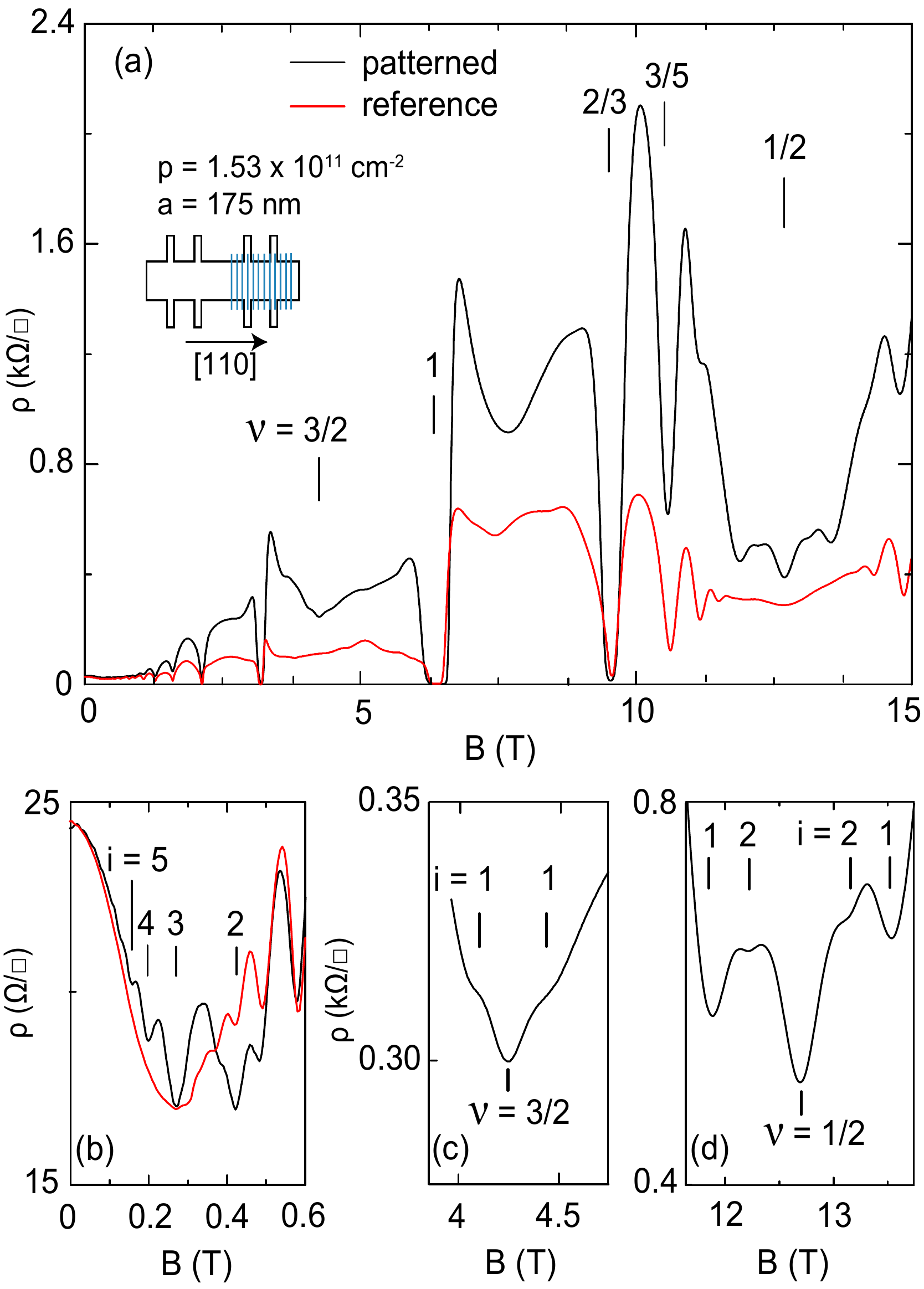}
\caption{\label{fig:Fig1} (color online) (a) Magnetotransport data at $T=0.3$ K from the patterned (upper trace) and reference (lower trace) regions of the $a$ = 175 nm sample at density of $p$ = $1.53 \times 10$ $^{11}$ cm$^{-2}$. The upper trace exhibits CF COs and a V-shaped dip around $\nu=1/2$ and $3/2$. (b) Low-field magnetoresistance of the patterned and reference regions. The patterned trace shows COs of holes. The values of the COs minima based on Eq.\;(\ref{eq:2}) are marked with indexed vertical lines. (c), (d) CF COs minima near $\nu=3/2$ and $1/2$ and their positions according to Eq.\;(\ref{eq:1}) marked with vertical lines ($i=1,2$).}
\end{figure}

Here we report magnetoresistance measurements in a high-mobility GaAs 2D $\textit{hole}$ system (2DHS) subjected to a strain-induced, unidirectional, periodic potential modulation. The data near $\nu=1/2$ exhibit clear signatures of ballistic transport of hole-flux CFs around $\nu=1/2$ (Fig.\;\ref{fig:Fig1}): a deep V-shaped resistance dip centered at $\nu=1/2$, followed by COs minima (labeled $i=1$ and 2 in Fig.\;\ref{fig:Fig1}(d)) and shoulders of higher resistance further away from $\nu=1/2$. The positions of the COs minima agree well with the commensurability condition for fully spin-polarized hole-flux CFs in a weak periodic $\textit{magnetic}$ field. Analysis of the envelope of the COs permits an estimation of the CF COs lifetime in a Dingle-factor-type plot and its direct comparison with CFs' quantum (Shubnikov-de Haas) and transport lifetimes.

Strain-induced lateral superlattice samples with different lattice periods were prepared from a 2DHS confined to a 175-\AA\--wide GaAs quantum well grown via molecular beam epitaxy on a (001) GaAs substrate. The quantum well is located 131 nm below the surface and is flanked on both sides by 95-nm-thick Al$_{0.24}$Ga$_{0.76}$As spacer layers and C $\delta$-doped layers. The 2DHS density at $T=0.3$ K is $p\simeq 1.5 \times 10^{11}$ cm$^{-2}$, and the mobility is $\mu=1.2\times10^{6}$ cm$^2$/Vs. We varied $p$ using an In back gate. Each sample had two Hall bars made by wet etching past the top doping layer. As illustrated in Fig.\;\ref{fig:Fig1}(a) inset, half of each Hall bar was covered with a grating of a high-resolution negative electron-beam (e-beam) resist, while the other half was kept unpatterned as a reference. The e-beam resist grating creates surface strain which in turn induces a periodic 2D density variation through the piezoelectric effect in GaAs \cite{Endo.PRB.63.113310.2001,Skuras.ARL.70.871.1997,Kamburov.PRB.2012}. The periods of the e-beam resist gratings were $a=100$, 150, 175, 200, 225, 250, and 300 nm. Measurements were made in a $^3$He refrigerator with base temperature of $\simeq$ 0.3 K using low-frequency lock-in techniques.

We first outline the theoretical formalism that describes CF COs before further presenting the experimental data. In the CF picture, the effective magnetic field $B^{*}$ felt by the CFs is determined by the external magnetic field $B$ and the CF density $n$ through the expression $B^{*}=B-B_{1/2}=B-2n\Phi_{0}$, where $B_{1/2}$ is the field at $\nu=1/2$, and $\Phi_{0}=h/e$ is the flux quantum \cite{Jain.PRL.63.199.1989,Halperin.PRB.47.7312.1993,Willett.PRL.71.3846.1993,Kang.PRL.71.3850.1993,Goldman.PRL.72.13.1994,Smet.PRL.77.2272.1996}. In light of this expression, local density variations induced by a potential modulation create strong local variations of $B^{*}$. Thus, a periodic density modulation leads to a periodicity of $B^{*}$ \cite{Peeters.PRB.47.1466.1993,Willett.PRL.78.4478.1997,Smet.PRB.56.3598.1997,Gerhardts.PRB.53.11064.1996,Smet.PRL.80.20.1998,Smet.PRL.83.2620.1999,Zwerschke.PRL.83.2616.1999,Oppen.PRL.80.4494.1998,Mirlin.PRL.81.1070.1998} and triggers 1/$B^{*}$-periodic oscillations of the magnetoresistance. The oscillations are quantitatively related to the commensurability between the CF cyclotron orbit diameter $2R_C^{*}$ and integer multiples of the modulation period. More precisely, the positions of resistance minima are given by the \textit{magnetic} commensurability condition \cite{Smet.PRL.80.20.1998,Smet.PRL.83.2620.1999}:
\begin{equation}\frac{2R_C^*}{a}=i+\frac{1}{4},\;\;\;(i=1,2,3,...),
\label{eq:1}
\end{equation}
where $R_C^*=2\hbar k_F^*/eB^*$ is the CF cyclotron radius; $k_F^*=\sqrt{4\pi p}$ is the CF Fermi wave vector and $p$ is the CF density which is equal to the 2D hole density ($p$) for oscillations near $\nu=1/2$. The above expression for $k_F^*$ assumes a circular Fermi contour and complete spin polarization of CFs, and is larger than its low-field (hole) counterpart by a factor of $\sqrt{2}$.

Concentrating on the $\nu=1/2$ case, as seen in Figs.\;\ref{fig:Fig1}(a) and (d), the magnetoresistance trace from a sample with a patterned region ($a=175$ nm) has prominent COs features near this filling. First- and second-order ($i=1,2$) CF COs appear on both sides of $\nu=1/2$ followed by shoulders of rapidly rising resistance \cite{Mirlin.PRL.81.1070.1998}. The positions of the $i=1,2$ COs minima based on Eq.\;(\ref{eq:1}) are marked with vertical lines in Fig.\;\ref{fig:Fig1}(d). A characteristic, deep, V-shaped resistance dip is also seen centered at $\nu=1/2$. It has been observed in previous experiments with electron samples and is attributed to channeled orbits \cite{Smet.PRL.80.20.1998,Smet.PRL.83.2620.1999} of CFs, similar to the resonant orbits of carriers occurring in modulated systems at low fields \cite{Beenakker.PRL.62.2020.1989,Vasilopoulos.PRL.63.2120.1989,Beton.PRB.42.9229.1990}.

In our data we observe magnetoresistance features in the vicinity of $\nu=3/2$ as well. They are highlighted in Fig.\;\ref{fig:Fig1}(c) along with vertical lines indicating the positions of the first ($i=1$) COs minima predicted by Eq.\;(\ref{eq:1}) for fully spin-polarized CFs near $\nu=3/2$. The positions of these minima, given explicitly in endnote 18 of Ref.\;\cite{Endo.PRB.63.113310.2001}, are slightly asymmetric with respect to the field position of $\nu=3/2$ because of the dependence of the CF density on magnetic field. Such asymmetry is observed in the data of Fig.\;\ref{eq:1}(c) as well.

The patterned region's trace also shows pronounced resistance minima near zero magnetic field (Fig.\;\ref{fig:Fig1}(b)) signalling COs of 2D holes \cite{Kamburov.PRB.2012}.  The field positions of the observed minima are consistent with the \textit{electrostatic} COs condition \cite{Gerhardts.PRB.53.11064.1996}:
\begin{equation}\frac{2R_C}{a}=i-\frac{1}{4},\;\;\;(i=1,2,3,...).
\label{eq:2}
\end{equation}
The vertical indexed lines shown in Fig.\;\ref{fig:Fig1}(b) mark the positions of these minima based on Eq. (2) and assuming a circular, spin-degenerate Fermi contour with $k_F=\sqrt{2\pi p}$. As detailed in Ref. \cite{Kamburov.PRB.2012}, the Fermi contours of the 2D holes in our samples are slightly warped; however, our COs data appear not to be sensitive to such warping and are consistent with circular contours. Note in Fig.\;\ref{fig:Fig1}(b) that the traces from both the patterned and reference regions exhibit Shubnikov-de Haas oscillations commencing at fields $B>0.3$ T. Both traces are also relatively flat near zero field. In particular, the reference region's pattern does not have the characteristic positive magnetoresistance typically seen when resonant orbits occur \cite{Beenakker.PRL.62.2020.1989,Vasilopoulos.PRL.63.2120.1989,Beton.PRB.42.9229.1990}, suggesting that the strain-induced potential modulation is too gentle for the Fig.\;\ref{fig:Fig1} sample to trap holes in resonant orbits. Some of our other samples, however, do exhibit the characteristic positive magnetoresistance at low fields \cite{Kamburov.PRB.2012}.

\begin{figure}[t]
\includegraphics[width=0.48\textwidth]{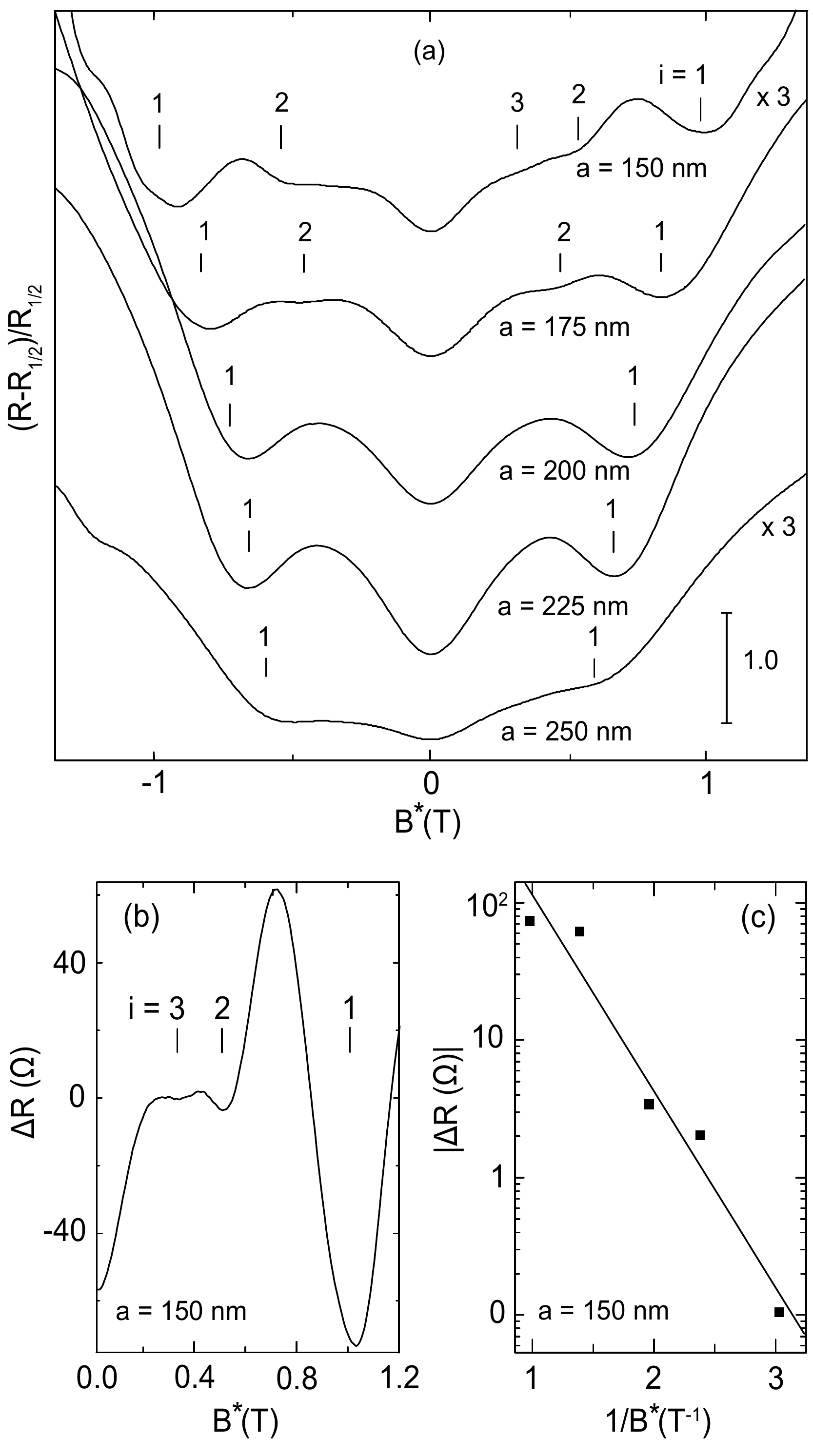}
\caption{\label{fig:Fig2} (a) Magnetotransport data near $\nu=1/2$ given in terms of the differential resistance, normalized to the value of resistance at $\nu=1/2$, are shown as a function of CF effective field ($B^{*}$). The traces are offset vertically without changing the scale. The top and bottom traces are multiplied by 3. The density of all samples is very close to $1.53 \times 10^{11}$ cm$^{-2}$. Vertical indexed lines mark the positions of the CF COs minima based on Eq.\;(\ref{eq:1}). (b) Oscillatory part of the magnetoresistance for the $a=150$ nm sample (see text). (c) Dingle plot of the maxima and minima of $|\Delta R|$ for the $a=150$ nm sample and a line fitted through the data points. }
\end{figure}

Returning to the $\nu=1/2$ data, in our experiments we find that the potential modulation period has an important effect on the appearance of the commensurability features near $\nu=1/2$. This dependence is illustrated in Fig.\;\ref{fig:Fig2} for five periods between 150 and 250 nm. Data from the $a=100$ nm and 300 nm period samples are not plotted here because the patterned regions of these samples do not show any sign of COs, and their magnetoresistance traces look very similar to those from the reference regions. In Fig. 2(a) the vertical lines mark the positions of the resistance minima in accordance with Eq.\;(\ref{eq:1}). For $B^{*}>0$, the expected positions of the COs minima match the observed positions well while for $B^{*}<0$ the observed minima are shifted towards $B^{*}=0$. A similar shift was also observed in electron samples and was attributed to the mixing of electrostatic COs \cite{Smet.PRL.80.20.1998}.

The dependence of the commensurability features on the modulation period seen in Fig. 2(a) is quite dramatic. For $a=250$ nm, the magnetoresistance is relatively flat and the features appear as weak minima. As $a$ is decreased, they become more pronounced and turn into deep minima. The V-shaped dip around $\nu=1/2$ also becomes stronger. As $a$ is further decreased, additional, second-order ($i=2$) features begin to develop and become well defined in the $a=175$ nm sample. In the $a=150$ nm trace, there is even a hint of a third-order ($i=3$) feature as well (also see Fig.\;\ref{fig:Fig2}(b)) but the amplitude of the COs envelope decreases. The $a=100$ and 300 nm samples (traces not shown) exhibit no COs at all.

The above dependence of the amplitude of the CF COs on $a$ can be qualitatively understood as follows. In the $a=200$ nm sample, the amplitude of the periodic surface strain is large enough to create a significant local density modulation which in turn creates a periodic $B^{*}$ and leads to strong COs features. As $a$ increases and becomes larger than the mean-free-path of CFs, the CFs have difficulty completing cyclotron orbits. The COs features become less pronounced, as seen in the bottom trace in Fig.\;\ref{fig:Fig2}(a), or disappear, as evidenced by the absence of COs in the $a=300$ nm sample \cite{footnote}. When $a$ is smaller than 200 nm, the mean-free-path of the CFs is long enough for CF cyclotron diameters to be commensurate with multiple times $a$. However, since the potential modulation created on the surface decays as $\propto exp(-2 \pi d/a)$, where $d$ is the 2DHS depth from the sample surface, the amplitude of the local density modulation is reduced \cite{Lu.APL.65.2320.1994,Endo.2005}. The reduction explains the smaller amplitude of the CF COs in the top trace in Fig.\;\ref{fig:Fig2}(a), and the absence of COs features for the $a=100$ nm sample.

The appearance of the higher order COs features in the $a=150$ nm sample allows for an analysis of the amplitude of the COs. From the amplitude of the differential resistance $(R-R_{1/2})/R_{1/2}$ given in Fig.\;\ref{fig:Fig2}(a), we obtain its oscillatory part $\Delta R$, shown in Fig.\;\ref{fig:Fig2}(b), by subtracting a straight line fit to the data in order to remove the positive background magnetoresistance. The maxima and minima of $|\Delta R|$ are plotted in Fig.\;\ref{fig:Fig2}(c) in a Dingle-type plot \cite{Dingle.1952}. Assuming $|\Delta R| \propto exp(-\pi/\omega^{*}_{^C} \tau^{_{CF}}_{^{CO}})$, the linear fit on the semi-logarithmic scale gives $\tau^{_{CF}}_{^{CO}}=16$ ps. Here $\omega^{*}_{^C}=eB^{*}/m^*$ is the CF cyclotron frequency, $e$ is the electron charge, and $m^*$ is the effective mass of CFs. We used $m^*=1.3m_e$, determined from our measurements of the energy gap for the $\nu=2/3$ fractional quantum Hall state in the reference region of the same sample ($m_e$ is the free electron mass). This value of $m^*$ is comparable to what was measured in previous studies of 2DHSs \cite{Manoharan.PRL.73.3270.1994}. The ballistic mean-free-path obtained from $\tau^{_{CF}}_{^{CO}}$ is $\lambda^{_{CF}}_{^{CO}}=0.2$ $\mu$m.

\begin{figure}[t]
\includegraphics[width=0.48\textwidth]{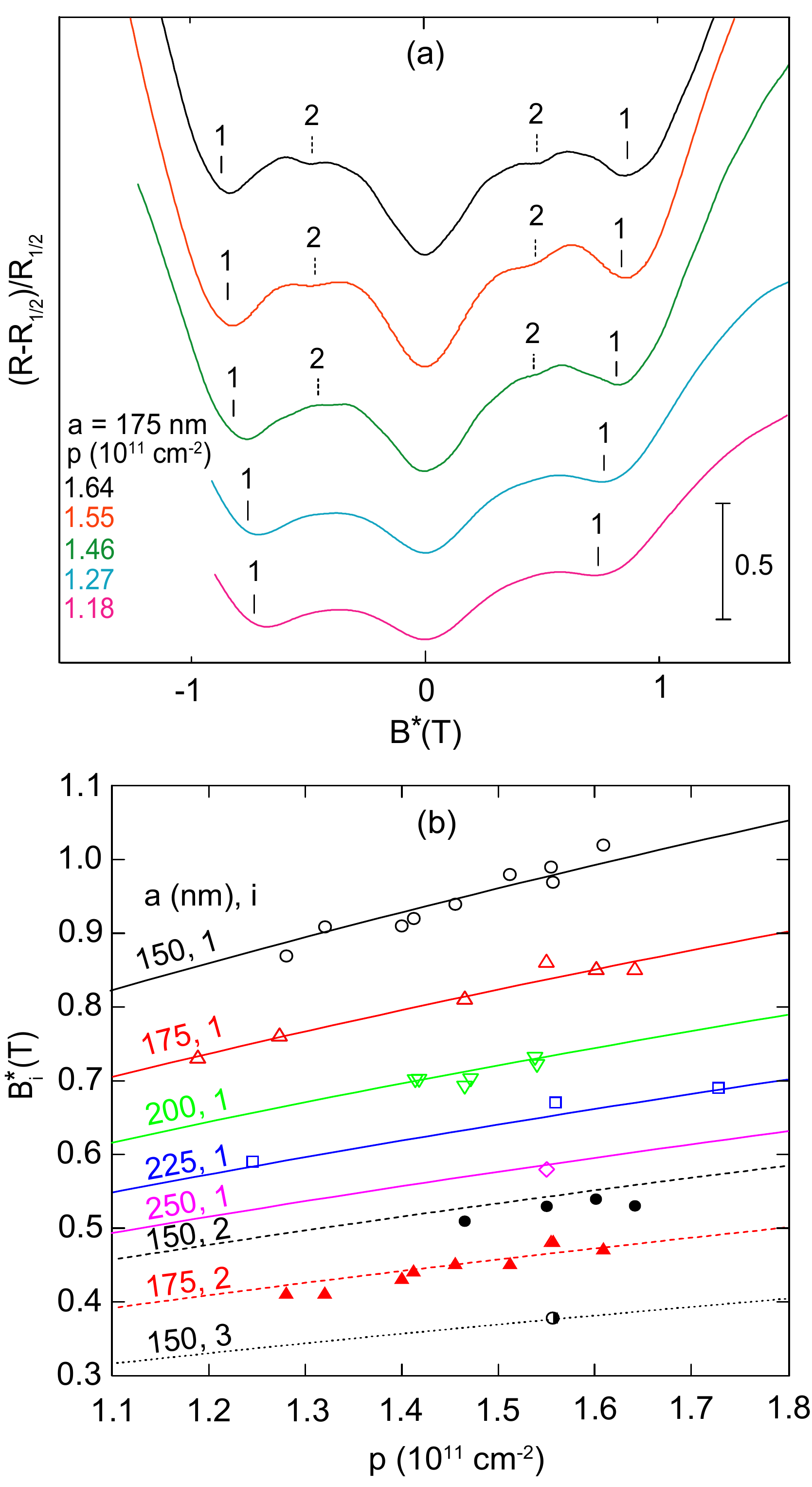}
\caption{\label{fig:Fig3} (color online) (a) Magnetotransport data of the $a=175$ nm sample as a function of density. Indexed lines correspond to the positions of the CF COs minima expected from Eq. (1). (b) Summary of the observed positions of the $B^*>0$ CF COs minima for different periods as a function of density. The lines give the predicted positions based on Eq.\;(\ref{eq:1}). }
\end{figure}

The value of $\tau^{_{CF}}_{^{CO}}$ found using this procedure is the first direct determination of the CF ballistic lifetime. It falls between the values for the CF Shubnikov-de Haas lifetime $\tau^{_{CF}}_{^{SdH}}=7.9$ ps we obtain from the $B^*$-dependence of the strength of fractional quantum Hall states in the vicinity of $\nu=1/2$ and the value of the CF transport lifetime $\tau^{_{CF}}_{tr}=63$ ps estimated from the resistivity at $\nu=1/2$. The finding that $\tau^{_{CF}}_{^{SdH}}<\tau^{_{CF}}_{^{CO}}<\tau^{_{CF}}_{tr}$ is similar to what has been reported for electrons and holes near $B=0$ \cite{Lu.1998,Lu.1999}, and attests to the qualitative similarity of CF transport to transport of electrons or holes. However, our result for $\tau^{_{CF}}_{tr}$ is much larger than the previously reported value of $7.5$ ps in Ref.\;\cite{Manoharan.PRL.73.3270.1994} for CFs in 2DHSs grown on (113)$A$ GaAs substrates, indicating that large-angle scattering is less prominent in our samples.

In Fig. 3(a) we present magnetoresistance traces from the $a=175$ nm sample as we vary the 2DHS density via biasing a back-gate electrode. The vertical lines mark the positions of the COs minima from Eq.\;(\ref{eq:1}). At low densities the $i=1$ minima are weak. As the density is increased, these minima become stronger, the V-shaped dip at $B^*=0$ deepens, and $i=2$ minima appear. Note that, in general, the positions of the $i=1$ minima are slightly asymmetric, with the $B^*<0$ minima appearing somewhat closer to $B^*=0$. The summary of the $B^*>0$ positions of the CF COs resistance minima for different periods and densities for all samples is shown in Fig.\;\ref{fig:Fig3}(b). The measured positions are given with symbols. Lines indicate the expected positions in accordance with Eq.\;(\ref{eq:1}). The data point for $i=3$ from the $a=150$ nm sample is from the trace in Fig.\;\ref{fig:Fig2}(b), and the one for the $a=250$ nm sample was obtained after the subtraction of a linear fit from the lower trace in Fig.\;\ref{fig:Fig1}(a). The agreement of the observed minima with those predicted by Eq.\;(\ref{eq:1}) is remarkable.

The data presented here demonstrate robust and strong CF commensurability features in 2D hole samples subjected to a strain-induced, unidirectional, periodic potential modulation. For appropriate modulation periods, we even observe higher order resistance oscillations near $\nu=1/2$ that correspond to CF cyclotron orbits spanning up to three times the superlattice period. An analysis of these oscillations allows us to extract a ballistic mean-free-path of about 0.2 $\mu$m for hole-flux CFs. The higher order CF COs minima we observe near $\nu=1/2$ are unprecedented as they have not been reported in 2D electron samples which typically have much higher mobilities. The exact circumstances which allow us to observe such strong CF COs features remain unclear but they are likely related to a combination of the strength, shape, and period of the potential modulation in our samples.

\begin{acknowledgments}

We acknowledge support through the DOE BES (DE-FG02-00-ER45841) for measurements, and the Moore and Keck Foundations and the NSF (ECCS-1001719, DMR-0904117, and MRSEC DMR-0819860) for sample fabrication and characterization. We thank A. Endo and Y. Iye for useful information and Tokuyama Corporation for supplying the negative e-beam resist TEBN-1 used to make the samples.

\end{acknowledgments}


\begin{thebibliography}{99}

\bibitem{Jain.Composite.Fermions} J. K. Jain, \textit{Composite Fermions} (Cambridge University Press, New York, 2007).
\bibitem{Jain.PRL.63.199.1989} J. K. Jain, Phys. Rev. Lett. \textbf{63}, 199 (1989).
\bibitem{Halperin.PRB.47.7312.1993} B. I. Halperin, Patrick A. Lee, and Nicholas Read, Phys. Rev. B \textbf{47}, 7312 (1993).
\bibitem{Willett.PRL.71.3846.1993} R. L. Willett, R. R. Ruel, K. W. West, and L. N. Pfeiffer, Phys. Rev. Lett. \textbf{71}, 3846 (1993).
\bibitem{Kang.PRL.71.3850.1993} W. Kang, H. L. Stormer, L. N. Pfeiffer, K. W. Baldwin, and K. W. West, Phys. Rev. Lett. \textbf{71}, 3850 (1993).
\bibitem{Goldman.PRL.72.13.1994} V. J. Goldman, B. Su, and J. K. Jain, Phys. Rev. Lett. \textbf{72}, 2065 (1994).
\bibitem{Smet.PRL.77.2272.1996} J. H. Smet, D. Weiss, R. H. Blick, G. Lutjering, K. von Klitzing, R. Fleischmann, R. Ketzmerick, T. Geisel, and G. Weimann, Phys. Rev. Lett. \textbf{77}, 2272 (1996).
\bibitem{Willett.PRL.78.4478.1997} R. L. Willett, K. W. West, and L. N. Pfeiffer, Phys. Rev. Lett. \textbf{78}, 4478 (1997).
\bibitem{Smet.PRB.56.3598.1997} J. H. Smet, D. Weiss, K. von Klitzing, P. T. Coleridge, Z. W. Wasilewski, R. Bergmann, H. Schweizer, and A. Scherer, Phys. Rev. B \textbf{56}, 3598 (1997).
\bibitem{Smet.PRL.80.20.1998} J. H. Smet, K. von Klitzing, D. Weiss, and W. Wegscheider, Phys. Rev. Lett. \textbf{80}, 4538 (1998).
\bibitem{Smet.PRL.83.2620.1999} J. H. Smet, S. Jobst, K. von Klitzing, D. Weiss, W. Wegscheider, and V. Umansky, Phys. Rev. Lett. \textbf{83}, 2620 (1999).
\bibitem{Peeters.PRB.47.1466.1993} F. M. Peeters and P. Vasilopous, Phys. Rev. B \textbf{47}, 1466 (1993).
\bibitem{Gerhardts.PRB.53.11064.1996} R. R. Gerhardts, Phys. Rev. B \textbf{53}, 11064 (1996).
\bibitem{Oppen.PRL.80.4494.1998} F. von Oppen, A. Stern, and B. I. Halperin, Phys. Rev. Lett. \textbf{80}, 4494 (1998).
\bibitem{Mirlin.PRL.81.1070.1998} A. D. Mirlin, P. Wolfle, Y. Levinson, and O. Entin-Wohlman, Phys. Rev. Lett. \textbf{81}, 1070 (1998).
\bibitem{Zwerschke.PRL.83.2616.1999} S. D. M. Zwerschke and R. R. Gerhardts, Phys. Rev. Lett. \textbf{83}, 2616 (1999).
\bibitem{Endo.PRB.63.113310.2001} A. Endo, M. Kawamura, S. Katsumoto, and Y. Iye, Phys. Rev. B \textbf{63}, 113310 (2001).
\bibitem{Skuras.ARL.70.871.1997} E. Skuras, A. R. Long, I. A. Larkin, J. H. Davies, and M. C. Holland, Appl. Phys. Lett. \textbf{70}, 871 (1997).
\bibitem{Kamburov.PRB.2012} D. Kamburov, H. Shapourian, M. Shayegan, L.N. Pfeiffer, K.W. West, K.W. Baldwin, and R. Winkler, Phys. Rev. B \textbf{85}, 121305(R) (2012).
\bibitem{Beenakker.PRL.62.2020.1989} C. W. J. Beenakker, Phys. Rev. Lett. \textbf{62}, 2020 (1989).
\bibitem{Vasilopoulos.PRL.63.2120.1989} P. Vasilopoulos and F. M. Peeters, Phys. Rev. Lett. \textbf{63}, 2120 (1989).
\bibitem{Beton.PRB.42.9229.1990} P. H. Beton, E. S. Alves, P. C. Main, L. Eaves, M. W. Dellow, M. Henini, O. H. Hughes, S. P. Beaumont, and C. D. W. Wilkinson, Phys. Rev. B \textbf{42}, 9229 (1990).
\bibitem{footnote} However, the absence of any commensurability features in the $a=300$ nm sample is puzzling, given that in the $a=175$ and 150 nm samples we observe resistance minima corresponding to $i \geq 2$ implying that in those samples the CFs complete cyclotron orbits with diameters that exceed 300 nm.
\bibitem{Lu.APL.65.2320.1994} J. P. Lu, X. Ying, and M. Shayegan, Appl. Phys. Lett. \textbf{65}, 2320 (1994).
\bibitem{Endo.2005} A. Endo and Y. Iye, J. Phys. Soc. Jpn. \textbf{74}, 1792 (2005).
\bibitem{Dingle.1952} R. B. Dingle, Proc. R. Soc. London A \textbf{211}, 517 (1952); E. N. Adams and T. D. Holstein, J. Phys. Chem. Solids \textbf{10}, 254 (1959); A. Isihara and L. Smrcka, J. Phys. C \textbf{19}, 6777 (1986).
\bibitem{Manoharan.PRL.73.3270.1994} H. C. Manoharan, M. Shayegan, and S. J. Klepper, Phys. Rev. Lett. \textbf{73}, 3270 (1994).
\bibitem{Lu.1998} J. P. Lu, J. B. Yau, S. P. Shukla, M. Shayegan, L. Wissinger, U. Rossler, and R. Winkler, Phys. Rev. Lett. \textbf{81}, 1282 (1998).
\bibitem{Lu.1999} J. P. Lu, M. Shayegan, L. Wissinger, U. Rossler, and R. Winkler, Phys. Rev. B \textbf{60}, 13776 (1999).

\end{thebibliography}
\end{document}